\documentstyle [twocolumn,epsf]{mn}
\newcommand{\mincir}{\raise
-3.truept\hbox{\rlap{\hbox{$\sim$}}\raise4.truept\hbox{$<$}\ }}
\newcommand{\magcir}{\raise
-3.truept\hbox{\rlap{\hbox{$\sim$}}\raise4.truept\hbox{$>$}\ }}
\newcommand{\minmag}{\raise
-3.truept\hbox{\rlap{\hbox{$<$}}\raise5.truept\hbox{$<$}\ }}
\newcommand{\be}{\begin{equation}}
\newcommand{\ee}{\end{equation}}
\newcommand{\ba}{\begin{eqnarray}}
\newcommand{\ea}{\end{eqnarray}}
\newcommand{\brr}{\begin{array}}
\newcommand{\err}{\end{array}}
\newcommand{\bc}{\begin{center}}
\newcommand{\ec}{\end{center}}

\title[PSCz Superclusters]
{PSCz Superclusters: Detection, Shapes \& Cosmological Implications}
\author[Basilakos, et al.]{S. Basilakos$^{1}$,
M. Plionis$^{2}$, M. Rowan-Robinson$^{1}$. \\
\vspace{0.1cm}
$^1$ Astrophysics Group, Imperial College London, Blackett Laboratory, 
Prince Consort Road, London SW7 2BW, UK\\
$^2$ Institute of Astronomy \& Astrophysics, National Observatory of Athens, 
Lofos Nimfon, Thesio, 18110 Athens, Greece \\
}

\begin{document}

\maketitle

\begin{abstract}
We study the possibility of correctly identifying, from the
smooth galaxy density field of the PSCz
flux limited catalogue, high density regions (superclusters) and
recovering their true shapes in the presence of a
 bias introduced by the coupling 
between the selection function and the constant 
radius smoothing.
We quantify such systematic biases in the smoothed PSCz density field 
and after applying the necessary corrections we study 
supercluster multiplicity and morphologies
using a differential geometry definition of shape. 
Our results strongly suggest that 
filamentariness is the dominant morphological feature
of PSCz superclusters. Finally, we compare our
results with those expected in three different cosmological models and find that
the $\Lambda$CDM model ($\Omega_{\Lambda}=1-\Omega_{m}=0.7$) performs 
better than $\Omega_{m}=1$ CDM models.

\end{abstract}

\vspace{0.2cm}

\section{Introduction}
The study of the distribution of matter on
large scales, based on redshift surveys of galaxies,
provides important constraints on models of cosmic structure formation.
It has been established that galaxy clusters
are not randomly distributed but tend to aggregate in
larger systems, the so called superclusters 
(cf. Bahcall 1988 and references therein). Superclusters 
are the largest, isolated and dynamically unrelaxed, due to their size, 
objects in the large scale distribution of matter and thus 
they are ideal probes of the initial conditions that gave rise to the
distribution of matter on such large-scales. One's hope is that they can be
used for cosmological studies in order to test theories of 
structure formation (cf. West 1989).  

Many authors have claimed that the large scale 
clustering pattern of galaxies is characterized by a filamentary 
and sheet-like distribution (cf. Zeldovich, Einasto, \& Shandarin 1982; 
Broadhurst et al. 1990; de Lapparent, Geller \& Huchra 1991).
West (1989), Plionis, Valdarnini \& Jing (1992) and Jaaniste et al (1997) 
studied the morphological properties (shape, size and orientation) of galaxy superclusters  
using Abell/ACO clusters and found, based on the simple principal axes method,
that the vast majority of the superclusters are flattened triaxial
objects, while Plionis et al. (1992) found a predominance of prolate 
(filament-like) superclusters. Further support to the filamentary 
case was presented recently by 
Sathyaprakash et al. (1998a), (1998b) analysing the shapes of the overdense 
regions using observational (IRAS 1.2Jy) and N-body simulation data.  

Studies based on the traditional indicators of clustering, like the 
two point correlation function, do not deepen our 
knowledge regarding the morphology and physics of structures on large scales.  
Therefore, many authors have been using different approaches, 
based on the geometrical properties of the large scale 
structure, in order to investigate such features. Indeed different methods like
minimal spanning trees, shape statistics (cf. Sahni \& Coles 1995
and references therein), 
genus-percolation statistics (Zeldovich et al. 1982; Gott, Mellot \&
Dickinson 1986; Coles \& Plionis 1991; 
Yess \& Shandarin 1996; Sahni, Sathyaprakash \& Shandarin
1997) and Minkowski functionals (cf. Mecke et al. 1994; 
Kerscher et al. 1997) 
have been used in order to describe the global geometrical and topological properties of 
the matter distribution utilising angular and redshift surveys of
galaxies as well as N-body simulations.
  
In this paper we use the recently completed PSCz-IRAS redshift survey 
in order: 
\begin{enumerate}
\item to investigate whether we can reliably identify
superclusters and measure their shapes in flux-limited galaxy samples, 
\item to measure the shape and size distribution of the PSCz superclusters and 
\item to investigate whether these distributions can be used as a cosmological probe.
\end{enumerate}
The plan of this paper is the following:
In section 2 we describe the PSCz data that we used, in Section
3 we present the supercluster identification procedure, tests of
systematic biases that enter in such a procedure and the detected PSCz
superclusters while in section 4 we present the supercluster shape
determination procedure, systematic effects that affect their shapes
and the PSCz supercluster shapes. In section 5 we compare
the PSCz results with the corresponding ones of three
cosmological models and finally in section 6 we draw our conclusions.

\section{The PSCz galaxy sample}
We use in our analysis the recently completed IRAS flux-limited 
60-$\mu$m redshift survey (PSCz) which is described in
Saunders et al. (2000). It is based on the IRAS Point Source 
Catalogue and contains $\sim 15500$ galaxies with flux $S_{lim}\ge 0.6$ Jy
covering the $84\%$ of the sky. 
The subsample we use, defined by a limiting galaxy 
distance of 240 $h^{-1}$ Mpc, contains $\sim 11823$ galaxies. 

\subsection{PSCz selection function}
Due to the fact that the PSCz catalogue is a flux-limited sample 
there is the well-known degradation of
sampling as a function of distance from the observer (codified by the
so called {\em selection function}).
Unless one corrects for this effect the derived galaxy density field will have
little to do with the true one. This is usually done
by weighting each galaxy by the inverse selection function,
$\phi^{-1}(r)$, assuming that the unobserved galaxies are spatially
correlated with the observed ones.
Note that according to the above the selection
function is defined as the fraction of the galaxy number density 
that is observed above the flux limit at some distance $r$. Therefore
\begin{equation}\label{eq:sf}
\phi(r)=\frac{1}{\langle \rho_{g} \rangle} \int_{L_{min}(r)}^{L_{max}} \Phi(L) dL
\end{equation} 
where $L_{min}(r)=4\pi r^{2} \nu S_{lim}$ is the luminosity of a source 
at distance $r$ corresponding to the flux limit $S_{lim}$, $\nu =$ 60-$\mu$m 
and $\langle \rho_{g} \rangle$ is the mean galaxy number density, given by 
integrating the luminosity function over the whole luminosity range, with
$L_{min}=10^{8} \; h^{2} L_{\odot}$ (since lower luminosity galaxies
are not represented well in the available samples; cf. Rowan-Robinson et al 
2000), and $L_{max}= 10^{13} \; 
h^{2} L_{\odot}$. Obviously, $\phi(r)$ is a 
decreasing function of distance because a smaller 
fraction of the luminosity function falls above the flux limit at greater 
distances. In this work we used 
a luminosity function of the form assumed by Saunders et al. (1990) with 
$L_{*}=10^{8.45} \; h^{2} L_{\odot}$, $\sigma=0.711$, $\alpha=1.09$ 
and $C=0.0308$ (cf. Rowan-Robinson et al. 2000). 

\subsection{PSCz galaxy distances from redshifts}
It is well known that the distribution
of galaxies in redshift space is a distorted representation of
that in real comoving space (Kaiser 1987).
The redshift distance of each galaxy can by found from:
\be
r=\frac{2 c}{H_{\circ}} \left(1-(1+z-\delta z)^{-1/2} \right)
(1+z-\delta z)^{3/2}
\ee
where $H_{\circ} = 100 \; h$ km s$^{-1}$ Mpc$^{-1}$ and $\delta z$ is a 
non-linear term to correct
the redshifts for the galaxy peculiar velocities:
\be \delta z =\frac{1}{c} ({\bf u}(r)-{\bf u}(0)) \cdot \hat{r} \ee
with ${\bf u}(0)$ the peculiar velocity of
the Local Group and ${\bf u}(r)$ the galaxy peculiar velocity 
at position ${\bf r}$, which can be found in linear theory by:
\be
{\bf u}(r)=\frac{H_{\circ} \beta}{4\pi} \int \delta_{\rm g}({\bf r}) 
\frac{ {\bf r}-{\bf r^{'}}}{\mid {\bf r}-{\bf r^{'}}\mid} {\rm d}^{3}{\bf r}
\ee  
where $\beta=\Omega^{0.6} b^{-1}$, assuming linear biasing
between the background matter and the galaxy fluctuation field $\delta_{\rm g}$.
Note that heliocentric redshifts are first transformed to the Local Group frame
using $c z \simeq c z_{\odot}+300 \sin(l)\cos(b)$ and then 
the distance to each galaxy is usually found by iteratively solving the above
set of equations. Such a procedure was used by Branchini et al (1999)
to recover the true density field of IRAS-PSCz galaxies in order to
study the galaxy density and velocity fields. In our work we
use their results (for $\beta=1$) which were kindly provided to us by
Dr. Branchini.

\subsection{Incomplete Sky Coverage}
In order to produce a continuous and whole sky density field it is
essential to treat the 16\% of the sky which is devoided of data 
(galactic plane, high cirrus emission areas and unobserved regions). 
We use the PSCz data of Branchini et al. (1999), in which they
followed the Yahil et al. (1991) method to fill the galactic plane region 
with synthetic objects that reproduce the mean density 
of galaxies in the regions residing nearby. At higher galactic
latitudes they fill the masked regions again with synthetic objects
so that they reproduce the overall mean number density of PSCz galaxies.

\section{Identifying Superclusters in the smooth PSCz density field}
\subsection{Smoothing Procedure}
For the purpose of this study we need to derive from the discrete
distribution of PSCz galaxies a smooth continuous density field.
To this end we use a Gaussian kernel on a $N^{3}$ grid:
\begin{equation}\label{eq:ker}
{\cal W}(|{\bf x}_{j}-{\bf x_g}|)  = \left( 2 \pi R_{{\rm sm}}^{2} 
\right)^{-3/2} \
\exp\left(-\frac{|{\bf x}_{j}-{\bf x_g}|^{2}}{2 R_{{\rm sm}}^{2}} \right).
\end{equation}
The smoothed density, at the grid--cell positions
$\bf{x_ g}$, is then:
\begin{equation}\label{eq:rhog}
\rho({\bf x_g}) = \frac{\sum_{j} \rho({\bf x}_{j}) 
{\cal W}(|{\bf x}_{j}-{\bf x_g}|)}
{\int {\cal W}(|{\bf x_{j}}-{\bf x_g}|) {\rm d}^{3}x},
\end{equation}
where the sum is over the distribution of galaxies with positions
${\bf x}_{j}$.

In our study we will use two smoothing radii, namely 
$R_{{\rm sm}} = 5 \; h^{-1}$ Mpc and 10 $h^{-1}$ Mpc, to probe 
essentially different supercluster sizes. We will extend our
analysis out to $r_{max}$, where $r_{max}=150$ and 240 $h^{-1}$ Mpc
respectively for the two smoothing radii while the size of each
cell is
set equal to $R_{{\rm sm}}$. Therefore we use a  grid of size of
$60^{3}$ and $48^{3}$ respectively.
Note also that $|{\bf x}_{j}-{\bf x_g}| \le 3 R_{{\rm sm}}$ and thus
the integral in the denominator of eq. (\ref{eq:rhog}) has a 
value $\simeq 0.97$.

In order to construct our supercluster candidates we select 
all cells with an overdensity above a chosen threshold and 
join together those having common boundaries.
Due to the fact that superclusters should be identified 
as very high density peaks,
the critical value of the overdensity threshold is defined
directly from the probability density function 
as its the 97$\%$-quantile value (using lower 
values we tend to percolate structures through the whole volume).

\subsection{Correcting Biases Introduced by Smoothing}
Due to the drop of the selection function as a function of
distance, shot-noise effects increase with distance.
Although Gazta\~naga \& Yokoyama (1993) have shown that the 
smoothing process itself considerably suppresses such shot--noise effects,
the coupling between the selection function and the constant 
radius smoothing will result in a distorted smoothed
density distribution, especially at large distances.
We expect that Gaussian spheres, centered on distant cells, will
overestimate the true density in regions where galaxies are
detected (due to the large $\phi(r)$ weighting), while in underdense regions 
they will underestimate the true density.

The decrease of resolution, due to the drop of the selection function
at large distances, can be dealt by increasing the smoothing radius as a
function of distance. Using the mean inter-galaxy separation as the
Gaussian smoothing radius is usually a good rule-of-thumb criterion
to suppress such effects in velocity field studies (cf. Branchini 
et al. 1999). However, it
cannot be used to identify and study 
supercluster shapes or in studies of the {\em pdf} and its moments, in which it is
necessary to have a constant resolution element through out the studied volume.

\subsubsection{Quantification of the effect}
We quantify this effect by using N-body simulations for which we have
the overall 3D ``galaxy'' distribution as well as the corresponding PSCz
look-alike distributions (a detail study of the effect and of the devised
correction procedure on the corresponding {\em pdf} statistics will be 
presented in Plionis, Basilakos \& Rowan-Robinson 2001). 
Here we just note that we use
a CHDM ($\Omega_{\rm{hot}}$=0.3 and $\Omega_{\rm{b}}=0.1$) 
simulation based on an optimize 
version of the Truncated Zeldovich
Approximation, described in detail in Borgani et
al. (1995; 1997). There are 256$^3$ grid points and as
many particles, while the simulation box has $L=480 \; h^{-1}$ Mpc
and periodic boundaries. The choice of the CHDM model was based solely
on the fact that
the corresponding distribution of particles that we tag as `galaxies'
have a correlation length that resembles that of the IRAS galaxies 
($r_{\circ} \simeq 4$ $h^{-1}$ Mpc). 

The main question that we want to answer now is: {\em How the `PSCz'-based
smoothed density field relates to the corresponding true underlying 3D density
field?}
We answer this question by comparing the smoothed 3D distribution of
the parent population of simulated ``IRAS'' galaxies with that of 
the corresponding PSCz
look-alikes, generated from this distribution.
We derive the statistical deviation of the smoothed `PSCz'-based 
distribution from the corresponding 3D one as a function of distance.
Therefore, we evaluate the following quantity:
\be
\Delta(r)=\frac{\rho_{\rm ``PSCz''}(r)-\rho_{\rm 3d}(r)}{\rho_{\rm 3d}(r)} \;\;.
\ee
If the selection process does not affect the `PSCz'-based density field 
then $\Delta(r)$ in the average should be negligible for all
distances. Indeed we find that $\langle \Delta(r) \rangle \simeq 0$ for
all distances (broken line in figure 1 \& 2). However, if we split the
densities in two subsamples: one higher and one lower than $\langle
\rho_{g} \rangle$, then we find a strong distant dependent effect (see
figures 1 \& 2), where the coupling between the selection function and the
constant radius smoothing results in an underestimation and
overestimation of the low and high density regions respectively, especially
at large distances.
\begin{figure}
\mbox{\epsfxsize=8.8cm \epsffile{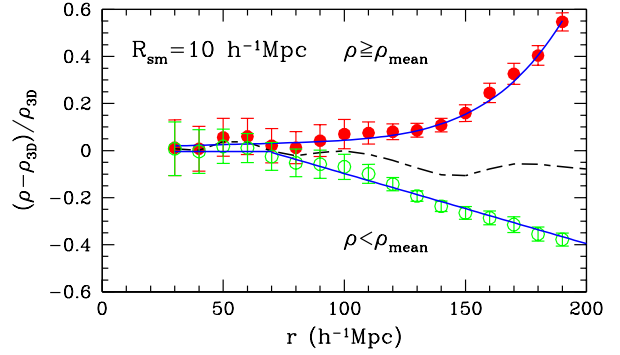}}
\caption{$\Delta(r)$ as a function of distance for the simulation
smooth density field with $R_{{\rm sm}}=10 \; h^{-1}$ Mpc. Filled
circles represent fluctuation values for $\rho> \langle \rho_{g} \rangle$
while open circles for $\rho < \langle \rho_{g} \rangle$.
The continuous lines
represents the best polynomial fit to $\Delta(r)$ and the broken line
represents the $\langle \Delta(r) \rangle$ values.} 
\end{figure}

\begin{figure}
\mbox{\epsfxsize=8.8cm \epsffile{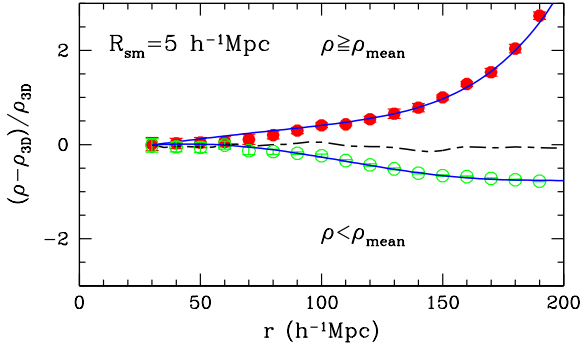}}
\caption{Same as for figure 1 but for the $R_{{\rm sm}}=5 \;h^{-1}$ Mpc case.}
\end{figure}
We can attempt to fit $\Delta(r)$ using a high-degree 
polynomial and then correct
the raw smoothed PSCz densities by weighting them with the factor 
$W(r)=(1+\Delta_{\rm fit}(r))^{-1}$. We have however to note that any
such multiplicative correction procedure cannot correct (artificial) 
zero densities. To this end we use a six degree polynomial 
and a $\chi^{2}$ minimization procedure to test the goodness-of-fit.
The fit, presented in figure 1 \& 2 as the continuous lines, has a reduced
$\chi^{2}$ value of $< 1$ indicating that the fit is good 
although the scatter is probably underestimated. In Plionis, Basilakos
\& Rowan-Robinson (2001) we will present details on the correction
procedure and the results in recovering the underlying {\em pdf} and
its moments. Here we test our procedure in recovering the true number,
size, multiplicity function and shapes of overdense regions (see below).

\subsubsection{Robustness of Supercluster Identification}
The above mentioned systematic effect should also affect the
connectedness of overdense regions and thus the detected
supercluster number,
size and shape. To this end we compare the number density of 
superclusters, found by our method in the smoothed 3D simulation
distribution with the average found in 10 `PSCz' realizations. 
As discussed before, we select cells that have an overdensity value
above a specific quantile of the {\em pdf}. However, since we are
interested in comparing the superclusters found in flux-limited samples
with the true underlying superclusters we have to make sure that we are
using equivalent thresholds. Such a choice is easy in the simulation test
case, in which we know {\em a priori} the true 3D {\em pdf}. 
We have found that in general a specific quantile of the raw
`PSCz' look-alike {\em pdf} is significantly larger than the
corresponding 3D {\em pdf}, an effect that 
our density correction procedure restores
to a good degree, especially in the {\em pdf} high
density tail (Plionis, Basilakos \& Rowan-Robinson 2001; in
preparation). Therefore, we use as our density threshold a 
specific {\em pdf} quantile.

\begin{figure}
\mbox{\epsfxsize=8.8cm \epsffile{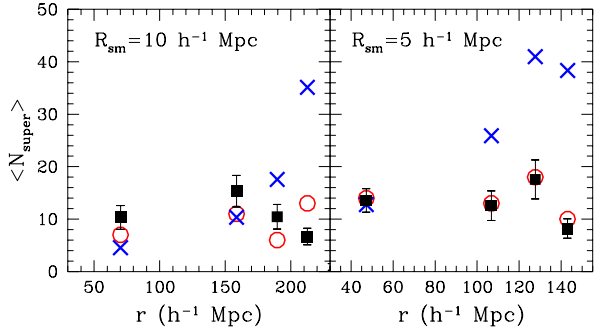}}
\caption{Supercluster number density in equal volume
shells as a function of distance for the 3D (open circles), raw-`PSCz'
(crosses) and corrected-`PSCz' (squares) cases.}
\end{figure}

In figure 3 we compare the supercluster number density in equal volume
shells ($\delta V\simeq 1.16 \times 10^{7} \; h^{-3}$ Mpc$^{3}$
for the $R_{sm}=10 \; h^{-1}$ Mpc case and 
$\delta V\simeq 2.83 \times 10^{6} \; h^{-3}$ Mpc$^{3}$ for the
$R_{sm}=5 \; h^{-1}$ Mpc case)
as a function of distance for the 3D, raw-`PSCz'
and corrected-`PSCz' distributions.
For the $R_{sm}=10$ $h^{-1}$ Mpc case we find that the
mean supercluster number density per equal volume
shell is 17$\pm$13, 10.9$\pm$3 and 9.3$\pm$3.3 for the `raw', 
`corrected' and true 3D density fields, while the corresponding
densities for the $R_{sm}=5$ $h^{-1}$ Mpc case are
30$\pm$13, 13$\pm$4 and 14$\pm$3.

It is evident that for the `raw-PSCz' case the
supercluster number density increases artificially with distance; 
at large distances the galaxy sampling rate is low 
and the smoothing procedure is unable to
produce a continuous density field, producing a large number 
of artificially disjoint regions.
This effect is efficiently suppressed once we correct
the smoothed density distribution according to our
correction procedure for both smoothing scales used. 

\begin{figure}
\mbox{\epsfxsize=8.8cm \epsffile{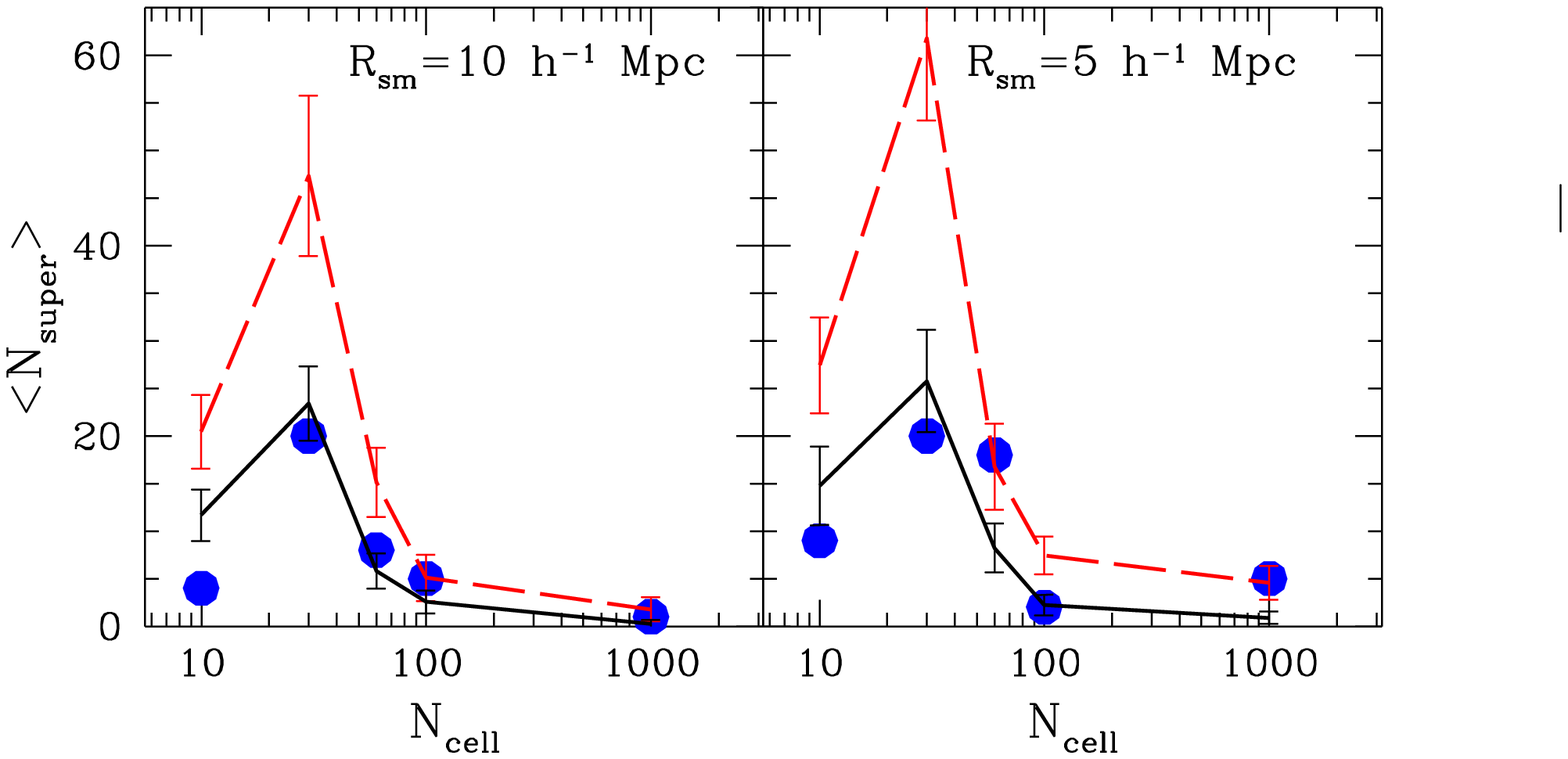}}
\caption{Supercluster multiplicity function for the 3D (points), raw-`PSCz'
(broken line) and corrected-`PSCz' (continuous line) cases.}
\end{figure}
In figure 4 we present the multiplicity function of the identified
superclusters. Again we see that our correction procedure is effective
in recovering the underlying multiplicity function.

We conclude that the connectedness of overdense regions
in flux limited
samples, with a selection function similar to that of IRAS 0.6 Jy galaxies,
can be recovered once we correct the `raw' smoothed density field
according to our procedure. We have verified, however, that
in order to restore such connectedness we need to correct the smoothed
density field limiting ourselves within $\sim 230$ $h^{-1}$ Mpc for the 
$R_{sm}\ge 10$ $h^{-1}$ Mpc case and within $\sim 150$ $h^{-1}$ Mpc for the 
$R_{sm}\ge 5$ $h^{-1}$ Mpc case. 

\subsection{Space Density and Multiplicity of PSCz Superclusters} 
In figure 5 we present the space density of PSCz superclusters in 
equal volume shells, having size as that in the previous section,
before and after correcting the density field
according to our procedure. We have also verified that 
variations of $\delta V$ do not alter our results.
\begin{figure}
\mbox{\epsfxsize=8.8cm \epsffile{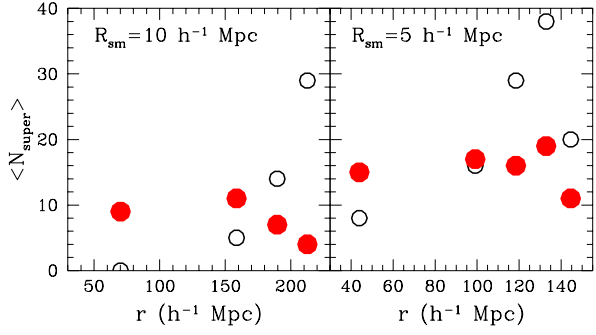}}
\caption{Mean number density of PSCz superclusters in equal volume shells,
based on the corrected (solid circles) and uncorrected (open circles)
smooth density fields.}
\end{figure}
For the $R_{sm}=10 \;h^{-1}$ Mpc case, in which we are revealing 
relatively large structures, we find in the uncorrected density
field that the first shell ($\mincir 130 \;h^{-1}$ Mpc), 
contains no supercluster, which evidently is false since
it is well known that within this distance there are several superclusters 
like the Great Attractor, Perseus-Pisces, Coma etc. 
We also observe the systematic increase of supercluster space density 
with distance in both $R_{sm}$ cases. However, once we correct the 
density field according to our procedure, discussed in 3.2, we 
recover a constant space density of superclusters in the different
equal volume shells. 

In figure 6, we present the PSCz supercluster multiplicity function,
for the uncorrected (dashed line) and corrected density fields
(continuous line), which will be used in the last section as a
cosmological probe.
\begin{figure}
\mbox{\epsfxsize=8.8cm \epsffile{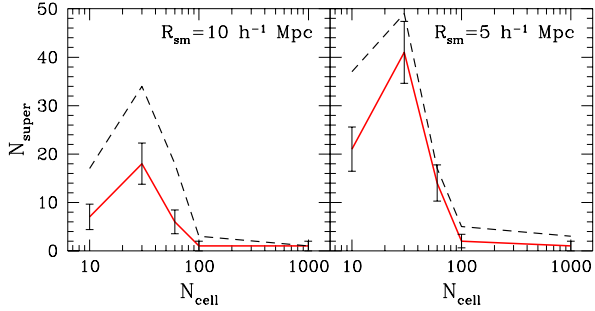}}
\caption{Multiplicity function of PSCz superclusters for the 
corrected (continuous line) and uncorrected (broken line)
smooth density fields.}
\end{figure}
In table 1 and 2 we present for both smoothing scales, the detected
PSCz superclusters, having a volume above the indicated threshold, as well as
their shape parameters (see section 4). We have
identified most of the known superclusters but also quite a few new
ones. Many other known superclusters, not shown in the tables, have been
found but fall below the specific volume threshold. 
\begin{table*}
\caption[]{Detected superclusters in the $R_{sm}=10$ $h^{-1}$ Mpc density
field with $V>2.5 \times 10^{4} \; h^{-3}$ Mpc$^{3}$ ($N_{cell} \ge
25$) and distance $\mincir 210 \; h^{-1}$ Mpc.}
\tabcolsep 7pt
\begin{tabular}{ccccccccl} 
Volume ($10^{4} \;h^{-3}$ Mpc$^{3}$) & $\langle d \rangle$ 
&$X_{sup}$ & $Y_{sup}$ & $Z_{sup}$ & $K_{1}$ & $K_{2}$ & $K_{1}/K_{2}$ & Name\\ 

        41&146.&  40.&-140.& -13.&  .095&  .113&  .839 &Pisces\\
        70&177.& -85.&-102.&-118.&  .104&  .076& 1.368 &\\
        36&195.&  10.&-101.&-167.&  .067&  .126&  .534 & \\
        30&148.&-107.& -96.&  35.&  .084&  .095&  .886 &\\
        38&201.&-187.& -75.& -10.&  .012&  .016&  .745 &\\
        27&152.& 104.& -55.& -95.&  .018&  .020&  .888 &\\
       103&171.&-157.&  58.&  38.&  .149&  .098& 1.510 & Far Shapley\\
        51&136.&-117.&  70.&  -4.&  .049&  .083&  .585 & Near Shapley \\
        33&110.& -32.&  75.&  74.&  .092&  .065& 1.415 & Hercules\\
        40&159.& -55.&  95.&-115.&  .043&  .066&  .645 & Near Horologium\\

\end{tabular}
\end{table*}

\begin{table*}
\caption[]{Detected superclusters in the $R_{sm}=5$ $h^{-1}$ Mpc density
field with $V>3.5 \times 10^{3} \; h^{-3}$ Mpc$^{3}$ ($N_{cell} \ge
28$) and distance $\mincir 140 \; h^{-1}$ Mpc.}
\tabcolsep 7pt
\begin{tabular}{ccccccccl} 
Volume ($10^{3} \;h^{-3}$ Mpc$^{3}$) & $\langle d \rangle$  
&$X_{sup}$ & $Y_{sup}$ & $Z_{sup}$ & $K_{1}$ & $K_{2}$ & $K_{1}/K_{2}$ & Name\\ 

      4.1&134.&  46.&-125.&   0.&  .054&  .118&  .458 & Pisces\\
      4.1&137.& -94.& -96.&  31.&  .044&  .058&  .764 & \\
      6.0&108.&  59.& -78.&  48.&  .057&  .033& 1.697 &\\
      6.0&105.&  32.& -58.& -81.&  .112&  .215&  .522 &\\
      3.5&123.&  52.& -46.& 102.&  .111&  .048& 2.328 &\\
      4.9& 89.&  72.& -47.& -24.&  .054&  .082&  .655 &\\
      5.0&128.&  26.& -36.&-120.&  .060&  .074&  .805 & Lepus\\
      5.4&138.&  95.& -46.& -90.&  .049&  .054&  .908 &\\
      7.0& 49.& -39.& -24.&  18.&  .078&  .113&  .692 & Pavo-Indus\\
      7.0& 49.&  46.& -14.&  -9.&  .116&  .053& 2.205 & Perseus-Pegasus\\
     15.1& 44.& -36.&  19.& -15.&  .080&  .055& 1.460 & Great Attractor\\
      6.3&137.&-118.&  49.&  50.&  .029&  .037&  .779 &\\
     11.8&132.&-114.&  67.&  -1.&  .071&  .158&  .445 & Near Shapley \\
      3.9&107.&  87.&  52.& -32.&  .000&  .000&  ---- & \\
      5.5&110.& -43.&  77.&  65.&  .055&  .138&  .400 & Hercules\\
      9.0&119.& -25.&  75.&  89.&  .090&  .196&  .459 &\\
      3.5& 78.&   3.&  77.&  14.&  .002&  .002&  .851 & Coma/A1367\\
      7.3&124.& -14.& 107.&  61.&  .063&  .106&  .591 &\\
      6.8&126.& -44.& 108.& -47.&  .033&  .063&  .517 &\\
      3.6&137.&   4.& 122.& -63.&  .045&  .108&  .418 & 
\end{tabular}
\end{table*}

\section{3d Supercluster Shapes}

\subsection{Shape Determination Method}
The supercluster shapes are defined by fitting ellipsoids to the 
data. Shapes are estimated for those ``superclusters'' that 
consist of 8 or more cells. To this end we use the moments of
inertia method (cf. Carter \& Metcalfe 1980; Plionis, Barrow \& Frenk
1991):
$I_{11}=\sum\ w_{i}x_{i}^{2}$,
$I_{22}=\sum\ w_{i}y_{i}^{2}$,
$I_{12}=I_{21}=\sum\ w_{i}x_{i}y_{i}$, 
$I_{13}=I_{31}=\sum\ w_{i}x_{i}z_{i}$, 
$I_{23}=I_{32}=\sum\ w_{i}y_{i}z_{i}$, 
with $w_{i}$ the statistical weight of each cell. This is defined
as the density fluctuation:
\begin{equation}
w_i = \frac{\rho_{i}(x_{gr})-\langle \rho_{g} \rangle }{\langle \rho_{g}
\rangle}
\end{equation}
where $\langle \rho_{g} \rangle$ is the mean galaxy number density.
Note that because of its symmetry we
diagonalize the inertia tensor
\begin{equation}\label{eq:diag}
{\rm det}(I_{ij}-\lambda^{2}M_{3})=0 \;\;\;\;\; {\rm (M_{3} \;is \; 
3 \times 3 \; unit \; matrix) } \;,
\end{equation}
obtaining the eigenvalues $I_{1}$, $I_{2}$, $I_{3}$ from which we define 
the shape of the configuration since the
eigenvalues are directly related to the three principal axes 
of the fitted ellipsoid.
The corresponding eigenvectors provide the direction of the principal axes.

\subsection{Geometrical Shape Statistics}
To identify the characteristic morphological features of superclusters 
and to investigate the geometrical properties of the large-scale
structure, we will use the differential geometry approach, introduced by
Sahni et al. (1998).  

We remind the reader the basic elements 
of differential geometry theory (cf. Lipschutz 1969). 
Assuming a local coordinate system on a surface with 
the integrating class ${\cal C}^{2}$, it is well known that the 
geometrical features of the surface are described by the first and second 
fundamental forms:

\begin{equation}
{\bf I}= {\bf dr} \cdot {\bf dr} = E {\rm d} \theta^{2}+2F 
{\rm d} \theta {\rm d} \phi+G {\rm d} \phi^{2}
\end{equation} 

\begin{equation}
{\bf II}= -{\bf dr} \cdot{\bf dn} = L {\rm d} \theta^{2}+ 
2M {\rm d} \theta {\rm d} \phi+N {\rm d} \phi^{2}
\end{equation} 
where $E={\bf r}_{\theta} \cdot {\bf r}_{\theta}$, 
$G={\bf r}_{\phi} \cdot {\bf r}_{\phi}$, 
$F={\bf r}_{\theta} \cdot {\bf r}_{\phi}$, 
$L={\bf r}_{\theta \theta} \cdot {\bf n}$, 
$G={\bf r}_{\phi \phi} \cdot {\bf n}$ and 
$M={\bf r}_{\theta \phi} \cdot {\bf n}$ 
\footnote{${\bf r}_{\alpha}=\frac{\partial {\bf r}}{\partial \alpha}$ and
${\bf r}_{\alpha \alpha}=\frac{\partial^{2} {\bf r}}{\partial \alpha^{2}}$}
(${\bf n}$ is the perpendicular vector 
${\bf n}=\frac{ {\bf r}_{\theta} \times {\bf r}_{\phi} }
{\mid {\bf r}_{\theta} \times {\bf r}_{\phi} \mid}$).  
Thus, the surface area, integrated mean curvature and genus can be written:
\be
S=\int \int \sqrt{EG-F^{2}} {\rm d}\theta {\rm d}\phi
\ee

\be
C=\int \int \frac{k_{1}+k_{2}}{2} {\rm d} S 
\ee

\be
{\cal G}=\frac{-1}{4\pi} \int \int k_{1}k_{2} {\rm d} S \;\; . 
\ee
where 
\be
k_{1}+k_{2}=\frac{EN+GL-2FM}{EG-F^{2}}
\ee 
and 
\be
k_{1}k_{2}=\frac{LN-M^{2}}{EG-F^{2}} \;\; .
\ee
The genus parameter characterises the surface topology;
multiply connected surfaces have ${\cal G}>0$ while simply connected 
ones have ${\cal G}<0$.
 
Furthermore, we define the 3 dimensional shape of structures
by fitting the best triaxial ellipsoid ($M=0$),
$${\bf r}(\theta,\phi)=I_{1} {\rm sin} \theta {\rm cos} \phi \hat{i}+
I_{2} {\rm sin} \theta {\rm sin} \phi \hat{j}+
I_{3} {\rm cos} \theta \hat{k} \;\;,$$ 
to the different isodensity contours having volume 
$V=\frac{4\pi}{3} I_{1}I_{2}I_{3}$ and $0\le \phi \le 2\pi$, $0\le \theta \le \pi$.   

Here we present the basic steps of the shape statistics 
that we will use, following the notation of Sahni et al. (1998). 
They introduced a set of three shapefinders
${\cal H}_{1}=V S^{-1}$, ${\cal H}_{2}=S C^{-1}$ and ${\cal H}_{3}=C$, 
having dimensions of length. In the same framework they defined two 
dimensional shapefinders $K=(K_{1},K_{2})$, where 
\be
K_{1}=\frac{ {\cal H}_{2}-{\cal H}_{1} }{ {\cal H}_{2}+{\cal H}_{1} } 
\ee
and
\be
K_{2}=\frac{ {\cal H}_{3}-{\cal H}_{2} } { {\cal H}_{3}+{\cal H}_{2} } \;\;, 
\ee
normalized to give ${\cal H}_{i}=R$ ($K_{1,2}=0$) for a sphere of radius $R$.
Therefore based on these shapefinders we can characterise the morphology of
cosmic structures according to the following categories:
\begin{itemize}
\item Pancakes for $K_{1}/K_{2}>1$
\item Filaments for $K_{1}/K_{2}<1$
\item Ribbons for $(K_{1},K_{2})=(\alpha,\alpha)$ with $\alpha \le 1$ and 
thus $K_{1}/K_{2} \simeq 1$.
\item Sphere for $I_{1} \simeq I_{2} \simeq I_{3}$ and thus $(K_{1},K_{2}) \simeq (0,0)$.
\item Ideal filament (1-d objects) for ${\cal H}_{1} \simeq 
{\cal H}_{2} << {\cal H}_{3}$ and thus $(K_{1},K_{2}) \simeq (0,1)$.
\item Ideal pancake (2-d object) 
for ${\cal H}_{1} << {\cal H}_{2} \simeq {\cal H}_{3}$ 
and thus $(K_{1},K_{2}) \simeq (1,0)$.
\end{itemize}
Finally, for the quasi-spherical objects the $K_{1,2}$ are small and
therefore their ratio ($K_{1}/K_{2}$) measures the deviation from spherical shapes. 
   
\subsection{Robustness of Supercluster Shape Determination}
Here we investigate how do the systematic effects, discussed in section 3.2,
affect the determination of supercluster
shapes. To this end we compare the shapes of the
superclusters, found by our method in the smoothed 3D simulation
distribution with those found in 10 'PSCz' realizations. 

\begin{figure}
\mbox{\epsfxsize=8.8cm \epsffile{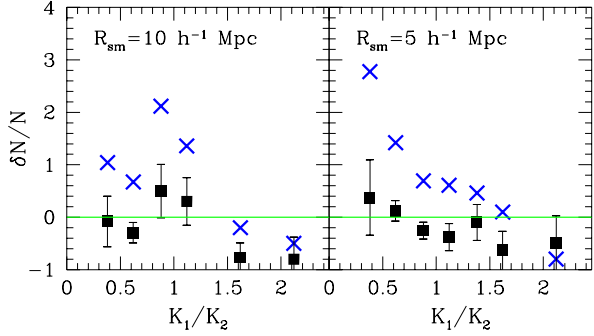}}
\caption{Shape Spectrum fluctuations
between the 3D and raw-`PSCz' case (crosses)
and between the 3D and the corrected-`PSCz' case (squares).}
\end{figure}
Performing many tests using different smoothing radii, we have found that 
the ratio of the dimensionless supercluster shapefinders (``shape spectrum'') 
is also affected from the bias discussed above.
Note that we exclude the ideal spherical surfaces, 
$I_{1} \simeq I_{2} \simeq I_{3}$, because their ratio $K_{1}/K_{2}$ 
tends to infinity. In figure 9 we present the fluctuations
of the shape spectrum between the 3D and raw-`PSCz' case 
and between the 3D and the corrected-`PSCz' case. 
Again it is evident that our
correction procedure suppresses significantly the systematic effects 
introduced by
the coupling between the selection function and the constant radius
smoothing for both smoothing radii.

We conclude that in order to correctly identify superclusters and their
connectedness, as well as their shapes from
flux limited samples, it is necessary to take into account the
bias of the smoothed density field, discussed in section 3.2,
and correct accordingly the smoothed galaxy distribution. 

\subsection{The PSCz Supercluster Shape Spectrum}
In figure 10 we present the uncorrected and corrected 
``shape spectrum'' for both smoothing radii.
It is obvious that the dominant feature of the PSCz 
supercluster shapes is filamentariness; ie., $K_{1}/K_{2}<1$ 
(in agreement 
with previous studies; see Sathyaprakash et al 1998). 

Furthermore, although the fraction of filamentary
superclusters ($K_{1}/K_{2}<1$) remain roughly the same
before and after the correction, there is a slight change of the
corrected `shape-spectrum' towards less extreme filamentary superclusters.
\begin{figure}
\mbox{\epsfxsize=8.8cm \epsffile{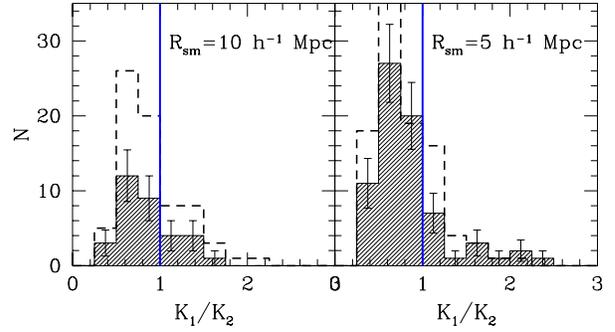}}
\caption{The corrected (hatched region) and uncorrected (dashed line) 
shape spectrum of PSCz superclusters for the two 
smoothing radii.}
\end{figure}
Regarding extreme shaped superclusters (with $N_{cell}\ge 8$), 
we have found, in the $R_{sm}=5 \; h^{-1}$ Mpc smoothed field, 
8 very filamentary superclusters with $K_{1}/K_{2} <0.45$ (among which
the Near Shapley and Hercules superclusters),
6 spherical and only 2 extreme pancake-like structures 
with $K_{1}/K_{2}>2$ (one
of which is the Perseus-Pegasus supercluster; see table 2). In the
$R_{sm}=10 \; h^{-1}$ Mpc smoothed field we find only one such
extreme filamentary supercluster.

\section{Comparison with Cosmological Models}
We use mock PSCz catalogues, generated from three large cosmological N-body
simulations, in order to investigate whether supercluster properties
(shapes and numbers) can discriminate between models. The cosmological
simulations are based on an AP$^{3}$M code and they are described 
in Cole et al. (1998). We consider three different 
cold dark matter models: (1) a flat low-density CDM model with 
$\Omega_{m}=1-\Omega_{\Lambda}=0.3$,
(2) a critical density universe $\Omega_{m}=1$ with power
spectrum $\Gamma=0.25$ ($\tau$CDM) and
(3) a critical density universe $\Omega_{m}=1$ with power 
spectrum $\Gamma=0.5$ (SCDM). The  
first two models are normalized by the observed cluster abundance at
zero redshift; having fluctuation amplitude in 8 $h^{-1}$Mpc scale of
$\sigma_{8}=0.55\Omega_{m}^{-0.6}$ (Eke, Cole, \& Frenk 1996),
while the third is COBE normalized with $\sigma_{8}=1.35$.

For each cosmological model we average results over
 10 nearly independent mock PSCz catalogues extending out to a 
radius of 170 $h^{-1}$Mpc (cf. Branchini et al. 1999). Good
care has been taken to center the catalogues to suitable LG-like observers
(having similar to the observed Local Group velocity, shear and overdensity).

We analyse the mock PSCz catalogues density fields (with $R_{sm}=5$ $h^{-1}$ Mpc)
in the same fashion as that of the observed PSCz catalogue 
and we compare the outcome of this procedure in
figure 11, where we plot the detected supercluster shape-spectrum and 
multiplicity function for all three models and PSCz data. 
Evidently the SCDM model performs worst in the comparison. 
In order to quantify the
differences between models and data we perform a standard 
$\chi^{2}$ test and we present the ${\cal P}_{>\chi^{2}}$
results in table 3. The only model that
is excluded by the data at a relatively high significance level 
is the SCDM model, while the
$\Lambda$CDM model best reproduces the observed supercluster 
shape-spectrum and multiplicity function. The $\tau$CDM is marginally
excluded by the multiplicity function comparison but only at a $\sim$
94\% level.

Comparing the models among themselves we see that the shape-spectrum is
insensitive to the different cosmologies, probably because supercluster
shapes reflect
the Gaussian nature of the initial conditions which are common to all
models. However, the supercluster multiplicity function is a strong 
discriminant between the models and therefore our results do
show a preference of the $\Lambda$CDM model over the 
$\Omega_{m}=1$ models that we have studied.

\begin{table}
\caption[]{$\chi^2$ probabilities (${\cal P}_{>\chi^{2}}$)
of consistency between data and
models as well as between the models themselves.}
\tabcolsep 8pt
\begin{tabular}{ccc}
Comparison Pair & Shape-Spectrum & Multiplicity fun. \\ 
PSCz - $\Lambda$CDM & 0.148 & 0.135 \\
PSCz - SCDM & 0.019 & 0.019 \\
PSCz - $\tau$CDM & 0.146 & 0.06 \\ \\
$\Lambda$CDM - SCDM & 0.71 & 5.2$\times 10^{-3}$ \\
$\Lambda$CDM - $\tau$CDM & 0.94 & 6.6$\times 10^{-4}$ \\
$\tau$CDM - SCDM & 0.07 & 2$\times 10^{-8}$ \\
\end{tabular}
\end{table}

\begin{figure}
\mbox{\epsfxsize=9cm \epsffile{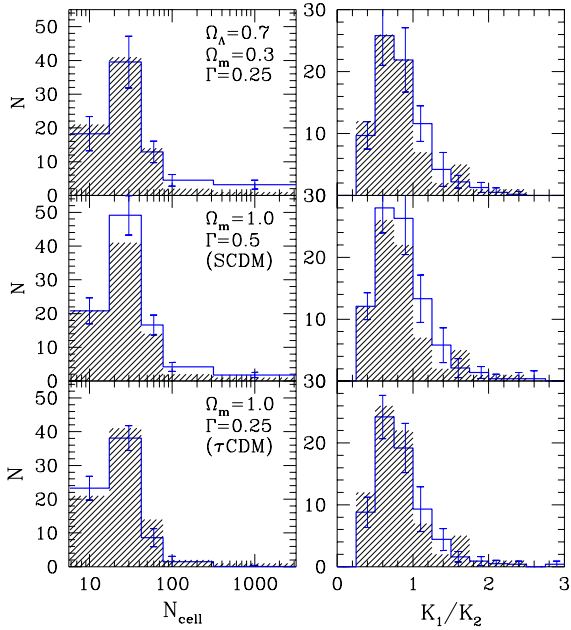}}
\caption{PSCz Superclusters comparison with Cosmological models: 
{\em Left panel}: Multiplicity function, {\em right panel}:
Shape spectrum. The PSCz results are represented by the hatched region.}
\end{figure}

\section{Conclusions}
We have studied the properties of superclusters detected in the 
smoothed PSCz galaxy density field as connected regions above
some overdensity threshold.
We have investigated, using simulations, the biases that enter
in the detection and shape determination of superclusters 
which are identified
in smoothed density fields of flux-limited galaxy samples.

We have devised a statistical approach to correct for such biases
which we have verified that indeed it recovers the underlying 3D
properties of superclusters.
To determine supercluster shapes we use the differential geometry 
approach of Sahni et al. (1998) and find that the dominant
supercluster morphological feature is filamentariness.

Finally, we have compared our PSCz supercluster results with the
corresponding ones generated from the analysis of three cosmological models (SCDM,
$\tau$-CDM and $\Lambda$CDM) and we find
that the model that best reproduce the observational results is the
$\Lambda$CDM model ($\Omega_{\Lambda}=0.7$).

\section* {Acknowledgements}
We thank Enzo Branchini for providing us with his reconstructed PSCz 
galaxy distribution and the PSCz mock catalogs and S. Coles for 
generating the simulations from which the mock catalogues 
have been extracted from. 
We also thank Andreas Efstathiou and Seb Oliver for their help in using IDL
to produce the 3D PSCz supercluster maps. M. Plionis acknowledges the 
hospitality of the Astrophysics Group of Imperial College, where 
a major part of this work was done.


{\small 
\begin{thebibliography}{}
\bibitem[]{}Bahcall, N. A., Ann. Rev. Astr. Ap., 26, 631
\bibitem[]{}Borgani, S., Plionis, M., Coles, P., Moscardini, L., 1995, MNRAS, 277, 1210
\bibitem[]{}Branchini, E., Plionis, M., 1996, ApJ, 460, 569
\bibitem[]{}Branchini, E., et al., 1999, MNRAS, 308, 1 
\bibitem[]{}Broadhurst, T. J., Ellis, R. S., Koo, D. C., Szalay, A. S., 1990, Nature, 343, 726
\bibitem[]{}Carter, D. \& Metcalfe, J., 1980, MNRAS, 191, 325
\bibitem[]{}Coles, P. \& Plionis, M., 1991, MNRAS, 250, 75
\bibitem[]{}Cole, S., Hatton, S., Weinberg, D. H., Frenk, C. S., 1998, MNRAS, 300, 945
\bibitem[]{}Davis, M., Efstathiou, G., Frenk, C. S.,White, S. D. M., 1985, ApJ, 292, 371
\bibitem[]{}de Lapparent, V., Geller, M. J., Huchra, J. P., 1991, ApJ, 369, 273
\bibitem[]{}Eke, V., Cole, S., Frenk, C. S., 1996, MNRAS, 282, 263
\bibitem[]{} Gott, J. R., Dickinson, M., Melott, A. L., 1986, ApJ, 306, 341
\bibitem[]{} Jaaniste, J., Einasto, M., Einasto, J., Andernach, H., Muller, V., 1997, A\&A, 329, 1
\bibitem[]{}Kaiser, N., 1984, ApJ, 284, L9
\bibitem[]{} Kerscher, M., Schmalzing, J., Retzlaff, J., Borgani, S., 
Buchert, T., Gottlober, S., Muller, V., Plionis, M., Wagner, H., 1997, MNRAS, 284, 73
\bibitem[]{}Lipschutz, M M., 1969, Theory and Problems of Differential Geometry, McGraw Hill.
\bibitem[]{} Mecke, K. R., Buchert, T., Wagner, H., 1994, A\&A, 288, 697
\bibitem[]{} Nolthenius, R., Klypin, A. Primack, J. R., 1994, ApJ, 1994, 422, 45
\bibitem[]{}Plionis, M., Barrow J.D., Frenk, C.S., 1991, MNRAS, 249, 662
\bibitem[]{}Plionis, M., Valdarnini, R., Jing, Y. P., 1992, ApJ, 398, 12
\bibitem[]{}Rowan-Robinson, M., et al., 2000, MNRAS, 314, 37 
\bibitem[]{}Sahni, V. \& Coles, 1995, Phys. REp., 262, 1
\bibitem[]{}Sahni, V., Sathyaprokash, B. S., Shandarin, S., 1998a, ApJ, 495, L5
\bibitem[]{}Sathyaprakash, S. B., Sahni, V., Shandarin, S., 1998b, ApJ, 508, 551
\bibitem[]{}Sathyaprakash, S. B., Sahni, V., Shandarin, S., Fisher, B. K., 1998, ApJ, 507, L109
\bibitem[]{}Saunders, W., Rowan-Robinson, M., Lawrence, A., Efstathiou, G., Kaizer, N., 
Ellis, R. S., Frenk, C. S., 1990, MNRAS, 242, 318
\bibitem[]{}Saunders, W., et al., 2000, MNRAS, {\em in press}, (astro-ph/0001117)
\bibitem[]{}Stompor, R., G\'orski, K. M., Banday, A. J., 1995, MNRAS, 277, 122 
\bibitem[]{}West, J. M., 1989, ApJ, 347, 610
\bibitem[]{}Yahil A., Strauss M., Davis M., Huchra J.P., 1991, ApJ,
372, 380
\bibitem[]{}Yess, C. \& Shandarin, S., 1996, ApJ, 465, 2
\bibitem[]{}Zeldovich, Ya. B., A\&A, 4, 84
\bibitem[]{}Zeldovich, Ya. B., Einasto, J., Shandarin, S., 1982, Nature, 300, 407
\end{thebibliography}
}
\end{document}